\documentclass[
reprint,
superscriptaddress,
 amsmath,amssymb,
 aps,
]{revtex4-1}

\usepackage[english]{babel}
\usepackage[T1]{fontenc}
\usepackage[utf8]{inputenc}
\usepackage{bm}
\usepackage{mathtools}
\usepackage[version=4]{mhchem}
\usepackage{nicefrac}
\usepackage{enumerate}
\usepackage[hidelinks]{hyperref}
\usepackage{color}
\usepackage{lipsum}

\hypersetup{
  colorlinks   = true, 
  urlcolor     = blue, 
  linkcolor    = blue, 
  citecolor   = red    
}

\definecolor{darkred}{rgb}{0.8, 0.0, 0.0}

\begin{document}

\title{Local embedding of Coupled Cluster theory\\into the Random Phase Approximation using plane-waves}
\author{Tobias Sch\"afer}
\email{tobias.schaefer@tuwien.ac.at}
\affiliation{Institute for Theoretical Physics, TU Wien, Wiedner Hauptstraße 8-10/136, A-1040 Vienna, Austria}
\author{Florian Libisch}
\affiliation{Institute for Theoretical Physics, TU Wien, Wiedner Hauptstraße 8-10/136, A-1040 Vienna, Austria}
\author{Georg Kresse}
\affiliation{University of Vienna, Faculty of Physics and Center for Computational Materials Science, Kolingasse 14-16, A-1090 Vienna, Austria}
\author{Andreas Gr\"uneis}
\affiliation{Institute for Theoretical Physics, TU Wien, Wiedner Hauptstraße 8-10/136, A-1040 Vienna, Austria}


\begin{abstract}

We present an embedding approach to treat local electron correlation effects in periodic environments.
In a single, consistent framework, our plane-wave based scheme embeds a local high-level correlation calculation (here Coupled Cluster Theory, CC),
employing localized orbitals, into a low-level correlation calculation (here the direct Random Phase Approximation, RPA).
This choice allows for an accurate and efficient treatment of long-range dispersion effects.
Accelerated convergence with respect to the local fragment size can be observed if the low-level and high-level long-range dispersion are quantitatively
similar, as is the case for CC in RPA.
To demonstrate the capabilities of the introduced embedding approach, we calculate adsorption energies of molecules on a surface and in a chabazite crystal cage,
as well as the formation energy of a lattice impurity in a solid at the level of highly accurate many-electron perturbation theories.
The absorption energy of a methane molecule in a zeolite chabazite, for instance, is converged with an error well below 20 meV
at the CC level.
As our largest periodic benchmark system, we apply our scheme to the adsorption of a water molecule on titania in a supercell containing more than 1000 electrons.

\end{abstract}

\maketitle


\emph{Introduction.} -- 
Embedding methods are ubiquitous in computational materials science and electronic structure
theory~\cite{Niranjan1999,Chulhai2018,Eskridge2019,Knizia2012,Bulik2014,Libisch2014}.
They typically partition a
system into a small fragment of interest that is treated with high
accuracy, and a surrounding environment treated with a less accurate
and thus numerically more affordable approach \cite{Inglesfield1981,Knizia2012,Niranjan1999,Chulhai2018,Eskridge2019,Knizia2012,Bulik2014,DoblhoffDier2018,Niemeyer2020,Manby2012,Goodpaster2014,Libisch2017,Lee2019,Sauer2019}.
Several embedding schemes have been proposed, ranging from point-charge embedding
\cite{Dick1958}, quantum-mechanics/molecular mechanics methods
\cite{Nobel2013}, density functional theory (DFT)-in-DFT approaches \cite{Nafziger2014,Wesolowski2015}
to ONIOM-type approaches that introduce several
layers, each of which may be treated by increasingly sophisticated
ab initio methods \cite{Vreven2003}.
Embedding methods have the potential
to treat local phenomena such as molecule-surface interactions or defects in solids with
high accuracy for system sizes that would otherwise be intractable.
In practice, a main challenge is to develop partitioning
schemes that achieve a smooth and rapid convergence of computed
properties with respect to the fragment size, enabling reliable and
numerically stable \emph{ab initio} predictions.


Here, we propose to embed a high level electronic structure theory,
the coupled cluster (CC) method \cite{shavitt2009many,Short.range.corCoeste1960,On.the.CorrelatCizek.1966}, into the direct random phase approximation
(RPA)~\cite{macke_uber_1950,pines_collective_1952,Kresse2009}.  This combination is an extremely promising approach from the
perspective of the many-electron perturbation series expansion,
because the RPA corresponds to a diagrammatic
many-electron theory that is included in coupled cluster single and
doubles (CCSD) theory \cite{Scuseria2008}, capturing the most important non-local
long-range electronic correlation effects, making a seamless
integration possible.  
An important advantage of the outlined
scheme is that all methods are implemented on the same footing, employing a plane
wave basis set that can be systematically improved by increasing a
single cut-off parameter.
In order to demonstrate the accuracy and the reduction of computational cost, we consider four benchmark systems.
As a simple and well-studied two dimensional material, we consider the physisorption of water on hexagonal boron nitride.
A realistic three dimensional surface adsorption is represented by water on titania.
Fully periodic bulk structures are covered by van der Waals bonded methane in a zeolite crystal and a sodium impurity in solid lithium chloride.
The atomic structures are summarized in the Supplemental Material \cite{suppl}.



\label{sec:algo}

\emph{Method and Implementation.} -- All correlation calculations employ a periodic Hartree--Fock (HF)
reference at the $\bm \Gamma$-point.
The HF calculations are performed using the Vienna ab initio simulation package
(\texttt{VASP})\cite{Efficiency.of.aKresse1996} and a plane wave basis set in the framework of the
projector augmented wave (PAW) method~\cite{Projector.augmeBlochl1994} (details on the used
PAW potentials are listed in \cite{suppl}).
The energy cutoffs for the plane wave basis set correspond to the largest default value of the employed
PAW pseudo potential of the involved elements (see \cite{suppl}). 
For the CCSD calculations, we have employed the \texttt{cc4s} code that has already been applied to study various ground state
properties of periodic systems~\cite{Gruber2018,Hummel2017}.
Our approach allows to compute correlation energies at different levels of theory based on the same set of electron repulsion integrals,
thus avoiding implementation-specific numerical issues.
For the embedding, we perform a simple partitioning of the occupied Hilbert space by selecting
Wannier orbitals obtained from a unitary transformation of the occupied Bloch orbitals. 
The Wannier orbitals allow to define a spatially localized subspace of the occupied Hilbert space.
The employed correlation method is restricted to excite only from this local subspace.
On the other hand, the virtual space is compressed using natural orbitals based on the reduced one-electron density matrix. This density matrix
is calculated considering excitations of the local
occupied subspace only into the complete virtual space, and it is calculated by a low-level correlation method. 
This partitioning enables to embed a high-level correlation method into a low-level correlation method. 
Denoting the local subspace as the fragment F and the rest of the supercell as R, the correlation energy using method M can be intuitively partitioned into
\begin{equation}
E^\text{M} = E^\text{M}_\text{F} + E^\text{M}_\text{R} + E^\text{M}_{\text{F}\leftrightarrow\text{R}} \;, \label{eq:EnPart}
\end{equation}
where $E^\text{M}$ is the correlation energy in the supercell, and the first and second term on the right hand side describe the correlation energy within the fragment F and the rest R, respectively.
These are uniquely defined by partitioning the occupied manifold into orbitals localized on the fragment and rest.
The contribution $E^\text{M}_{\text{F}\leftrightarrow\text{R}}$ is hence also unambiguously defined and essentially captures all
correlation effects between the fragment and the rest. We now assume
that the  correlation energy between the fragment and the rest can be accurately described by the low level correlation method.
Consequently, we define the correlation energy obtained by embedding correlation method $\text M_1$ (e.g. CCSD) into method $\text M_2$ (e.g. RPA) by
\begin{align}
E^{\text{M}_1\text{:}\text{M}_2} &= E^{\text{M}_1}_\text{F} + E^{\text{M}_2}_\text{R} + E^{\text{M}_2}_{\text{F}\leftrightarrow\text{R}} \;,  \nonumber\\
&=  E^{\text{M}_1}_\text{F} +  E^{\text{M}_2} - E^{\text{M}_2}_\text{F}  \;. \label{eq:emb}
\end{align}
The second line follows from the identity $E^{\text{M}_2}_\text{R} + E^{\text{M}_2}_{\text{F}\leftrightarrow\text{R}} = E^{\text{M}_2} - E^{\text{M}_2}_\text{F}$ and is used in practice, since it contains well defined
contributions only, either in the entire supercell $E^{\text{M}}$ or in the fragment $E^{\text{M}}_\text{F}$.
As already alluded to above, partitioning is here based on {\em partitioning of the occupied manifold}.
The correlation energy contribution to the adsorption energy of a molecule on a surface, for instance,
can then be calculated as
\begin{equation}
E_\text{ads} = E^{\text{M}_1\text{:}\text{M}_2}_\text{int} - E^{\text{M}_1\text{:}\text{M}_2}_\text{far} \;, \label{eq:molSurface}
\end{equation}
where int and far stand for the adsorbed and distant situation of the molecule, respectively.

\begin{figure}
\includegraphics[width=0.5\textwidth]{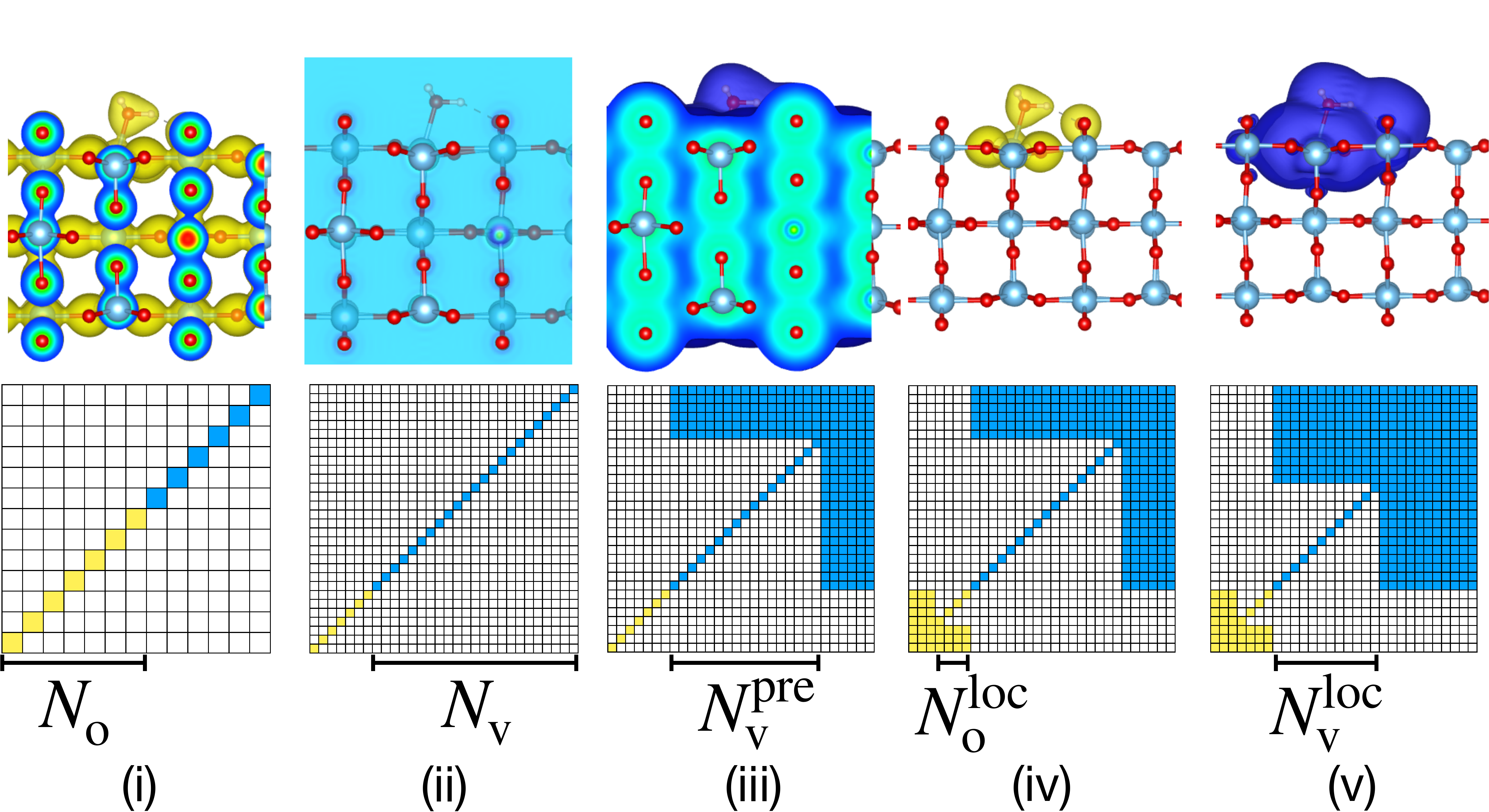}
\caption{
Illustration of the algorithm steps (i)--(v) used to obtain localized occupied and virtual orbitals for a
fragment centered at the adsorbed water molecule on the titania surface.
Isosurfaces of the charge densities computed from the full ($N_{\text o}$) and localized ($N_{\text o}^{\text loc}$) occupied
orbital manifolds are depicted in (i) and (iv), respectively.
(ii), (iii) and (v) show isosurfaces for charge densities computed from the (natural) virtual orbital manifolds.
The structure of the Fock matrix after recanonicalization is depicted schematically.
The non-zero matrix elements for occupied and unoccupied orbital indices are colored yellow and blue, respectively.
} 
\label{fig:method}
\end{figure}

We now explain the individual technical steps required to calculate the correlation energy, $E^\text{M}_\text{F}$, of a fragment F using method M.
The charge densities of the employed occupied and unoccupied orbital manifolds as well as the structure of the Fock matrix
in steps (i)-(v) as described below are depicted in Fig.~\ref{fig:method} for the case of a water molecule on titania.

\begin{enumerate}[(i)]

\item \emph{Mean-field ground state}\\
Self-consistent solution of the mean-field equation of the entire supercell, obtaining $N_{\text o}$ occupied Bloch orbitals.

\item \emph{Unoccupied space}\\
Diagonalize the mean-field Hamiltonian in the given basis to obtain all $N_{\text v}$ unoccupied Bloch orbitals. Here, $N_{\text o}+N_{\text v}$ is equal to the number of plane-wave
basis functions controlled by the energy cutoff.

\item \emph{First compression of unoccupied space}\\
Calculate the natural orbitals at the level of the RPA  \cite{Ramberger2019} for the entire periodic system in order to precompress
the unoccupied space to $N^\text{pre}_v = X \cdot N_{\text o}$ orbitals (usually $X=40,...,60$), as described in the Supplemental Material \cite{suppl}.
We note that natural orbitals yield correlation energies, which converge quickly to the complete basis set limit \cite{Gruneis11,Ramberger2019}.

\item \emph{Define fragment}\\
Transform all $N_{\text o}$ Bloch orbitals to $N_{\text o}$ Wannier orbitals.
To determine the Wannier functions, we use a projection based scheme, where all
occupied orbitals are rotated using a unitary matrix such that the overlap with a set of
$N_{\text o}$ local trial functions is maximized.
As trial functions, we choose atom-centered pseudo partial waves given in the PAW pseudopotential,
however, also simple Slater-type functions or atomic orbitals could be used.
A detailed description can be found in Sec. IV. B. in Ref. \cite{Engel2020}.
The fragment is defined by selecting atoms, based on the radial distance from the adsorbed molecule or the lattice impurity.
The Wannier orbitals associated with these atoms thus form the basis of the subspace of the occupied space. 
Finally, we diagonalize the mean-field Hamiltonian in the basis of the $N^\text{loc}_\text{o}$ selected Wannier orbitals (re-canonicalization).

\item \emph{Second compression of unoccupied space}\\
Calculate the natural orbitals of the fragment only, based on an approximated MP2 scheme
\cite{Gruneis11} using $N^\text{pre}_v$ pre-compressed orbitals to further compress and localize the unoccupied space to $N^\text{loc}_{\text v} = Y \cdot N^\text{loc}_{\text o}$ orbitals ($Y\leq X$, usually $Y=40,...,60$), as described in the Supplemental Material \cite{suppl}. In this step, excitations are only
allowed from the selected $N^\text{loc}_\text{o}$ orbitals localized on the fragment. As a matter of fact, one could 
again use the RPA density matrix, however, the approximate MP2 scheme is computationally more efficient at this stage.

\item \emph{Local correlation energy}\\
Decompose the electron repulsion integrals (ERI) into auxiliary three index quantities using $N_{\text F}$ auxiliary field variables, i.e. auxiliary basis functions
(for details, see Supplemental Material \cite{suppl} and Ref. \cite{Hummel2017}). Calculate the correlation energy
(MP2, RPA, CCSD, ...) for $N^\text{loc}_\text{o}$ local occupied and $N^\text{loc}_\text{v}$ unoccupied orbitals and employing $N_{\text F}$ auxiliary field variables.
We stress that despite using a plane wave basis set,
all dimensions that contribute to the scaling of the computational complexity of the electron correlation methods exhibit a linear scaling with fragment size:
$N_{\text F} \propto N^\text{loc}_{\text v} \propto N^\text{loc}_\text{o} $. 
\end{enumerate}

Note, that the employed Wannier orbitals are by no means maximally localized.
However, the resulting subspaces are well defined and allow for a systematic convergence of the local correlation energy with respect to the extent of the fragment. 

We explore the performance of different embedding approaches:  CCSD:RPA, CCSD:dMP2, and MP2:dMP2.
Here, dMP2 refers to the direct MP2 contribution only (without the exchange-like MP2 contribution), while RPA always stands for direct particle-hole RPA.
From the computational point of view, embedding in RPA and dMP2 is particularly favorable due to existing low-scaling
implementations, $\mathcal{O}(N^3)$, for plane-waves \cite{Kaltak2014,Schafer2016}.
This allows to study systems with up to a thousand occupied
and hundred thousand virtual bands / plane-waves on current supercomputers.
We stress that all correlation energy approximations used in the present work employ HF orbitals and eigenvalues.
We also show results for interaction energies calculated from the bare fragment energies without embedding corrections,
i.e. $E^{\text{M}_1}_\text{F}$ instead of $E^{\text{M}_1\text{:}\text{M}_2}$, see Eq. (\ref{eq:EnPart}).


\emph{Results.} -- In this work we aim to reproduce correlation energies for a specific well defined plane wave basis set.
Reference values of the entire supercells at the RPA and MP2 level are calculated using approximation-free low-scaling implementations in {\sc vasp} \cite{Kaltak2014,Schafer2016}.
Comparison with experimental results would require additional considerations, for instance, basis set convergence corrections, finite size and thermal corrections.
We now present the results of our embedding scheme. Each system has a discussion part followed by technical details.

\begin{figure*}
\centering
\includegraphics[width=1.0\textwidth]{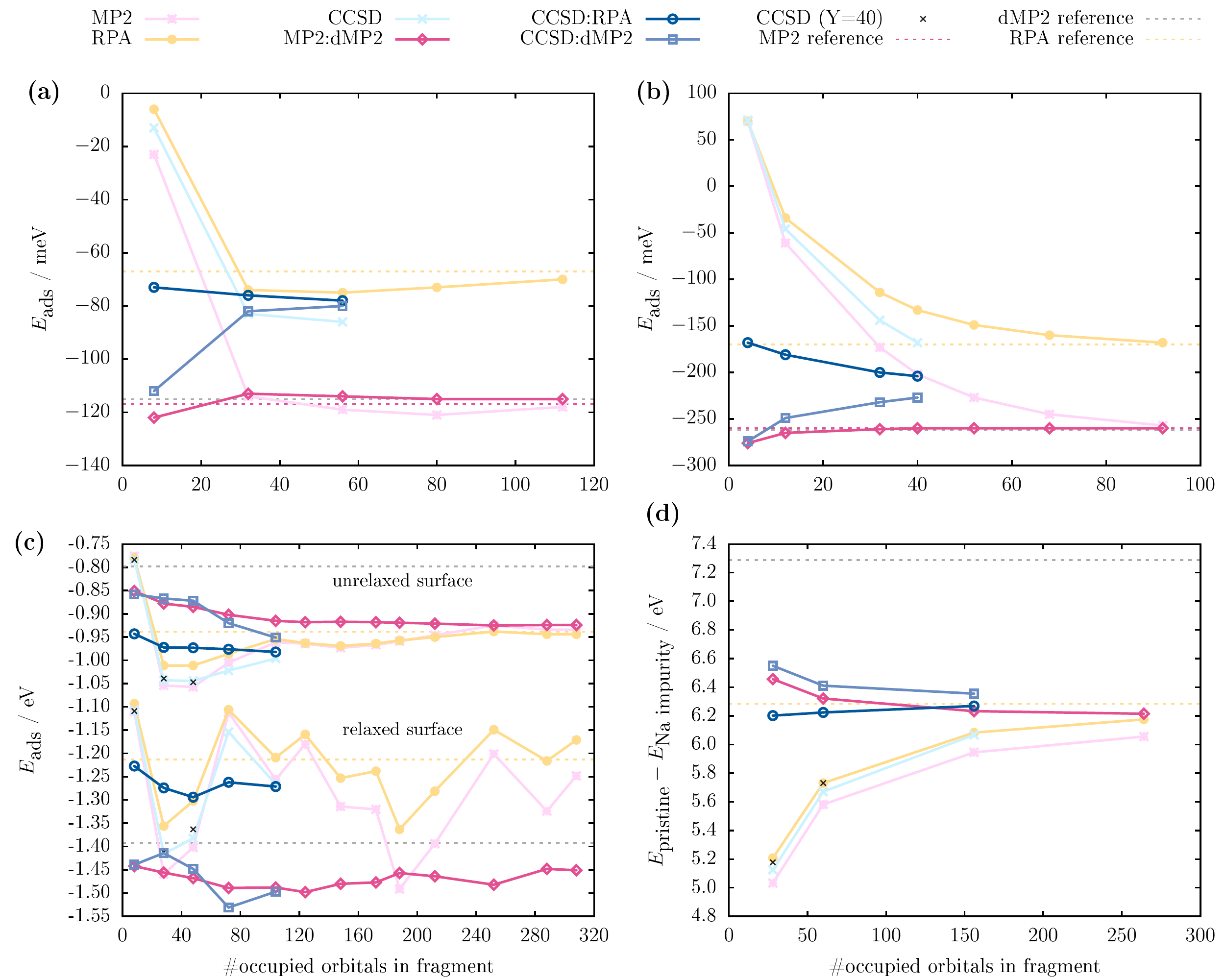}
\caption{Convergence of the adsorption and formation energies of the considered benchmark systems with respect to the fragment size: (a) water on hexagonal boron nitride, (b) methane in a chabazite crystal cage, (c) water on titania, (d) Na impurity in LiCl. The data points include the Hartree-Fock energy for the entire supercells. For more details we refer to the results section. The horizontal axes refer to the number of occupied spatial orbitals in the local fragment. }
\label{fig:results}
\end{figure*}

\smallskip

\noindent\emph{Water on $h$-BN.}
Figure \ref{fig:results} (a) shows the convergence of the electron correlation contribution to the adsorption energy
of water on the large band gap material $h$-BN. $h$-BN was modeled by a single two-dimensional $6\times 6$ layer. 
Clearly, even the bare fragments (without embedding into a low-level method) converge fairly rapidly to values close to the periodic reference case. 
For MP2:dMP2, an almost flat line is obtained, quite clearly improving on the bare MP2 fragment results.
The CCSD:RPA and CCSD:dMP2 results both clearly converge to the same final value, with the CCSD:RPA error being
only $15\,\text{meV}$ even for the smallest fragment. Here, CCSD:dMP2 converges from below,
since dMP2 overestimates long-range correlation effects, something observed for other materials as well.
\emph{Technical details:}
The supercell contains $72+3$ atoms and $144+4$ valence bands, where the second number corresponds to the molecule.
The HF contribution to the binding energy is $36\,\text{meV}$.
As trial functions to construct the fragments, we use the s and p partial waves of the PAW potential of the nitrogen atoms only.
Here, we use a large basis set, defined by a plane wave cutoff energy of $500 \,\text{eV}$. 
The virtual space is compressed to $40$ natural orbitals per occupied orbital, i.e. $X=60$ in step (iii), $Y=40$ in step (v).

\smallskip

\noindent\emph{Methane in zeolite crystal.}
Absorbing a methane molecule in a zeolite crystal cage (\ce{AlHO24Si11}) leads to fragments extending in all three spatial dimensions.
Again, we observe a rapid convergence of the CCSD:RPA curve, see Fig. \ref{fig:results} (b).
Again, an almost flat line is obtained for MP2, if embedded in dMP2.
CCSD:RPA and MP2:dMP2 follow an exponential decay, indicating that the local MP2/CCSD corrections collect only exchange like contributions, while long-range
correlations are covered by the corresponding low-level method. 
This is true for MP2:dMP2 by definition, but less obvious for CCSD:RPA.
The bare correlation energies converge smoothly but slowly.
This slow convergence is due to the slow but steady accumulation of long-range correlation contributions, revealing the necessity to
embed the high-level method into a low-level method with a similar long-range correlation description.
Here, CCSD:dMP2 seemingly also exhibits fast convergence, but the decay is not simply exponential.
\emph{Technical details:}
We note, that the supercell contains $37+5$ atoms and $96+4$ valence bands.
The HF contribution to the binding energy is $51\,\text{meV}$.
As trial functions to construct the fragments, we use the s and p partial waves of the PAW potential of the oxygen atoms only.
The virtual space is compressed to $60$ natural orbitals per occupied orbital, i.e. $X=60$ in step (iii), $Y=60$ in step (v).

\smallskip

\noindent\emph{Water on titania.}
A more realistic surface adsorption is considered with water on titania (101). 
Here, 3 layers of \ce{TiO2} are used to model the surface.
Furthermore, we compare the convergence with fragment size for an unrelaxed and relaxed surface.
The relaxed surface is obtained using DFT-PBE and clearly illustrates that the small displacements caused by the relaxation 
of the slab can cause a very slow non-monotonic convergence of the differences in the correlation energy.  
While the bare correlation energies show a slow but recognizable convergence on the unrelaxed substrate, there is no clear convergence visible for the relaxed structure, see Fig. \ref{fig:results} (c).
Note, that for this system, the choice of the low-level method is crucial.
Since the fragments 1 to 3 involve only occupied orbitals in the first layer, the dispersion interaction with deeper layers is left to the low-level method.
This separates the CCSD:RPA and CCSD:dMP2 curves by about $100\,\text{meV}$ ($200 \,\text{meV}$) for the unrelaxed (relaxed) surface for the first three fragments.
For the unrelaxed surface, this mismatch is significantly reduced for fragments 4 and 5. 
These include atoms from the second \ce{TiO2} layer.
For the relaxed surface, even larger fragments are necessary to resolve the mismatch between CCSD:RPA and CCSD:dMP2.
We note, however, that the bare CCSD results fluctuate around CCSD:RPA, while CCSD:dMP2 exhibits a large offset.
Furthermore, in both cases (relaxed and unrelaxed) CCSD:RPA exhibits a significantly smoother behavior compared to CCSD:dMP2. 
We estimate the remaining error of CCSD:RPA to be less than $40\,\text{meV}$ for the unrelaxed situation.
For the relaxed substrate, the CCSD:RPA curve appears to be converged to less than $80\,\text{meV}$, however, further investigations
are indispensable to validate this estimation. 
\emph{Technical details:}
Here, the supercell contains $144+3$ atoms and $576+4$ valence bands.
The HF binding energy contribution is $-744\,\text{meV}$ for the unrelaxed and $-979\,\text{meV}$ for the relaxed situation.
As trial functions to construct the fragments, we use the s and p partial waves of the PAW potential of the oxygen and titanium atoms.
The virtual space is compressed to $40$ natural orbitals per occupied orbital, i.e. $X=60$ in step (iii), $Y=40$ in step (v).
We note that in this case the CCSD energies for 40 virtual orbitals per occupied orbital have been approximated
using a basis set correction computed on the level of MP2 such that:
$E^{\text{CCSD}+\Delta \text{MP2}}_{(Y=40)}= E^{\text{CCSD}}_{(Y=10)}-E^{\text{MP2}}_{(Y=10)}+E^{\text{MP2}}_{(Y=40)}$
The good agreement between $E^{\text{CCSD}+\Delta \text{MP2}}_{(Y=40)}$
and $E^{\text{CCSD}}_{(Y=40)}$ for the smaller fragments as shown by the CCSD results and black crosses in Fig.~\ref{fig:results}~(c)
justifies this approximation, which is necessary to make the CCSD calculations for the larger fragments feasible.


\smallskip

\noindent\emph{Na impurity in LiCl.}
The formation of a lattice impurity is another widely studied local physical phenomenon.
For simplicity, we limit ourselves to the calculation of the energy difference between the pristine crystal and the defect structure, where one lithium atom is replaced by a sodium atom. 
The defect structure is relaxed with DFT-PBE in order to incorporate lattice distortions.
CCSD:RPA and MP2:dMP2 converge exponentially to $6.32$ and $6.22\,\text{eV}$, respectively, see Fig. \ref{fig:results} (d).
According to this extrapolation, the third fragment of CCSD:RPA ($6.27\,\text{eV}$) is converged with an accuracy of $0.05 \,\text{eV}$. 
The decay of CCSD:dMP2 is not simply exponential, but seemingly provides an equally high accuracy at fragment three ($6.36\,\text{eV}$).
For the CCSD energies (calculated with $Y=10$) we used the same basis set correction, as described in the discussion of water on titania.
\emph{Technical details:}
The supercell contains $216$ atoms and $436$ (432) valence bands, where the number in brackets corresponds to the pristine cell.
The HF contribution is $1.47\,\text{meV}$.
As trial functions to construct the fragments, we use the s and p partial waves of the PAW potential of the chlorine atoms (and additionally of the single sodium atom for the doped structure).
The virtual space is compressed to $40$ natural orbitals per occupied orbital, i.e. $X=60$ in step (iii), $Y=40$ in step (v).

We note that while this manuscript was finalized, a related study using a similar embedding approach was published as pre-print~\cite{lau2020}.
However, there are a number of crucial differences between this work and Ref.~\cite{lau2020} that we would like
to emphasize in the following. Ref.~\cite{lau2020} employs atom-centered basis functions to calculate adsorption energies on surfaces of mostly two dimensional sheets.
The authors employ MP2 to correct for finite size errors of coupled cluster energies.
This MP2 finite size correction is computationally more expensive than the proposed RPA and dMP2 correction that formally scale only as $\mathcal{O}(N^3)$ with respect to the system size.
Furthermore, while the RPA is well defined for metals, MP2 yields divergent correlation energies in metallic systems, which limits its scope as embedding method to insulators.


\emph{Summary and Conclusions.} --  We have presented a computationally efficient embedding scheme for coupled cluster theory calculations
of local phenomena in solids and on surfaces. We have applied the proposed method to an impurity in the bulk as well as molecular adsorption
problems in a chabazite crystal and on substrates with different dimensionalities.
All calculations have been performed using a plane wave basis set, demonstrating a seamless integration between periodic RPA methods
and wavefunction based coupled cluster theory calculations.
Our findings demonstrate that embedding coupled cluster theory in the RPA constitutes a stable and numerically efficient
scheme that yields rapidly converging results with respect to the local fragment size, even for three dimensional systems
with significant atomic relaxation effects.
In most materials, CCSD:dMP2 and CCSD:RPA approach the CCSD limit from opposite directions.
This allows for a useful consistency check and a concise  error estimation. \\
Furthermore our findings show that higher dimensional systems are in general less amenable to embedding methods, requiring larger fragment sizes to achieve converged results.
Yet, recent methodological advancements make it possible to treat sufficiently large
fragments with coupled cluster theory such that the desired level of high precision and accuracy can be obtained.

\section*{Supplementary Material}
See the supplementary material for details of the used PAW pseudopotentials, a description of the scheme to compress the unoccupied space, a description of the computation of the optimized auxiliary fields, and the used atomic structures (\texttt{POSCAR} files).

\section*{Acknowledgements}
A.G. and T.S. thankfully acknowledge support and funding from the European
Research Council (ERC) under the European Unions
Horizon 2020 research and innovation program (Grant Agreement No 715594).
F.L. acknowledges COST action CA18234.
The computational results presented have been achieved in part using
the Vienna Scientific Cluster (VSC).

\section*{Data availability}
The data that supports the findings of this study are available within the article and its supplementary material \cite{suppl}.



\bibliography{paper}

\end{document}